# Immersive Serious Games for Learning Physics Concepts: The Case of Density

Iuliia Zhurakovskaia[1], Jeanne Vézien[1], Cécile de Hosson[2], and Patrick Bourdot[1]

[1] Université Paris-Saclay, CNRS, LISN, 91400 Orsay
[2] Université de Paris, LDAR, 75205 Paris

**Abstract.** Training students in basic concepts of physics, such as the ones related to mass, volume, or density, is much more complicated than just stating the underlying definitions and laws. One of the reasons for this is that most students have deeply rooted delusions and misconceptions about the behavior of objects, sometimes close to magical thinking. Many innovative and promising technologies, in particular Virtual Reality (VR), can be used to enhance student learning. We compared the effectiveness of a serious immersive game in teaching the concept of density in various conditions: a 2D version in an embedded web browser and a 3D immersive game in VR. We also developed a specific questionnaire to assess students' knowledge improvement. Primary results have shown an increase in learning efficiency using VR. Also, most students were able to see the shortcomings of their initial theories and revise them, which means that they improved their understanding of this topic.

**Keywords:** Density, Virtual Reality, Science Education.

## 1   Introduction

Density can be derived as a unitless number, being the ratio of the volumetric mass of a (homogenous) body over the one of a reference object (usually water at 4°C, of density equal to 1g/cm3).

In order to teach density, Smith [1] mentioned that it is important to show early on the correct connection between density and flotation (an object denser than water will sink while it will float if less dense). If the learned materials do not fit the students' intuitive framework, they tend to distort the learning to accommodate their beliefs. Also, they can assimilate the new material as a separate system without any relations to the real world. Consequently, they "learn" the material as it is without understanding it, will pass exams, but fail to get a clear understanding of the underlying concept [1].

Strike and Posner [2] argued it is important to *show* students that their current concepts are wrong and do not stand up to serious scrutiny, and based on this, provide them an experimental path to the correct framework.

Recent STEM education research [3] reported that students who used VR for education and students using a regular desktop version exhibited similar results. Moreover, the results of immersion (3D) and desktop (2D) learning gave better results than actual field trips [3]. Based on these findings, and because current sanitary conditions encouraged remote teaching, we decided to evaluate teaching the concept of density in two conditions: in 2D (remotely: keyboard/mouse interaction on a 2D web-based simulation) and in 3D (in person: immersive VR experiment using a run-of-the-mill HMD).

Following a didactic approach, the main tasks we have identified are: (i) to provide a good understanding of different aspects of density for students; analyze and assess this understanding; (ii) to clarify the distinction between density, weight, and volume in students' understanding; (iii) to examine the effectiveness of the serious game approach in teaching density and compare the results with traditional didactic approaches.

This study leverages three separate tools: VR, Serious Games, and Didactics. VR technology allows generating a previously unavailable experience by interacting directly with simulated content that reproduces existing physical phenomena of all sorts. Serious games, also known as game-based learning, consists of any game that aims to enhance the player's knowledge. Such games use pedagogy to influence the learning experience [7][8]. The purpose of didactic research is to study the questions raised by teaching and the acquisition of knowledge in various scholarly disciplines.

One should note that, at this point, VR has never been used to teach the concept of density. Because this classic concept is addressed early in most teaching curricula, we selected it as a good candidate to evaluate the efficiency of VR to comprehend physics in a classroom setting.

## 2   Online Questionnaire

To establish a reliable baseline of knowledge regarding the subject of density, we first designed a dedicated online questionnaire based on [4]. The questionnaire contains 13 basics questions about the density concept, each with a 4-step confidence level.

Analysis of the a priori responses to the questionnaire showed that most of the students shared identical misconceptions on the subject (e.g., most students have not integrated the idea that the density of a solid is formally defined in relation to water) and that the knowledge regarding density was largely incomplete.

## 3   2D game

The second part of the research consisted in creating a 2D game designed as a web browser experience (Fig. 1). Participants used a regular screen (a laptop or PC) and mouse interaction to play within a 2D simulation, organized as a game with several "levels". This game was developed on the Unity engine [6].

The game is decomposed into several stages: a pre-test evaluates the current knowledge of students, followed by a training session introducing the main objects and possible interactions to familiarize the user with the environment of the game, followed by three different game stages, described below. A bonus level completes the game for extra fun, followed by a post-test questionnaire, which consists of the same questions as the pre-test but in another order.

3We designed three sequential gaming conditions: Condition 1 (C1) - All objects have the same volume but not the same mass; Condition 2 (C2) - All objects are with the same mass but not the same volume; Condition 3 (C3) - Two tanks filled with different liquids (water and oil) and two sets of the objects, same ones as in C1. The bonus level is a simple version of C3 (no game score, no predictions, quicksilver instead of oil).

All cubes have individual characteristics: size (volume), texture (dots, with varying spacing), and weight (mass). The texture is indicative of the density: the more dots, the higher the density [1].

The main task consisted of placing cubes in a tank full of liquid (water); the user then had to make a prediction about the objects' behavior in the water before observing the experimental outcome. Also, C1 introduced the Roberval Balance as a metaphor to compare cubes according to their mass.

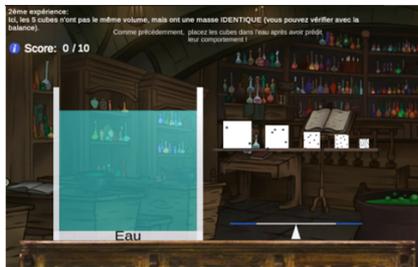 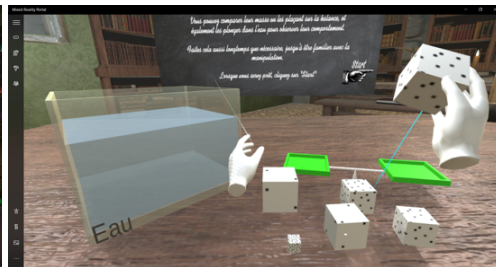

**Fig. 1.** 2D game.  **Fig. 2.** 3D game.

## 4      Immersive 3D game

The third part of the experiment consisted in using the same game as in the 2D part, but this time interacting within an immersive VR environment (Fig. 2).

Affordable price and reasonable quality of immersion hardware and software were paramount to this project. We selected the Lenovo Explorer VR headset with two controllers and the Unity game engine as an adequate setup.

All the interaction techniques exploited in this setup are based on the common "simple virtual hand" metaphor. Cubes could be picked up with both virtual hands, manipulated freely, and dropped (not thrown).

One of the main differences between 2D and 3D games was the possibility to manage the actual gaming conditions around participants. Aside from dimension and interaction modes, the main difference between the two versions of the game was the availability of sound. We didn't use audio in the 2D game because we could not control the audio conditions in the users' environment (some users could have muted the sound). In contrast, this was possible in the 3D VR on-site experiment. To provide more immersion and convey additional information (e.g. mass) about objects, each participant was equipped with headphones and binaural sound rendering was provided.

Representing weight in VR is a major problem because although force-feedback devices do exist, such haptic interactors are very expensive, cumbersome, and tend to be fragile. Using sounds to convey mass information to students seemed, therefore, a promising alternative approach [5].

**Experimental Procedure**. The local ethical committee approved the following experiment, and proper sanitary procedures to prevent Covid-19 contamination were applied [9]. For better immersion and to avoid additional device manipulations, users passed the pre-test in the virtual environment.

Then the students underwent a training session. This initial phase consisted of presenting the setup in a sandbox "no-task" configuration. General instructions were given on how to grasp cubes and place them in different locations.

During experiments, participants read all instructions on a virtual blackboard placed in front of them. In each trial, they could manipulate the cubes at will, with no time constraint. Participants performed a sorting task and, when a tank was present, made a prediction (correct answer + 2 points, wrong answer -1 point) about the behavior of the cube in the liquid (sink, stay in the middle, float), before they could observe the actual behavior. The game stages were identical to those of the 2D game.

The final post-test was a copy of the pre-test but with a reordering of the questions.

Following the VR experiment, participants were asked to undergo a 15-minutes semi-directed interview. The objective was to collect some data relative to subjective perceptions regarding the general quality of the environment and the interaction, the task itself, perceived object affordance, perception of weight, and overall satisfaction with the experiment.

## 5 Results

### 5.1 Online Questionnaire

After analyzing the online questionnaire, we exploited the R factor to structure the profiles of our respondents' answers. R computes the similarity of a set of answers and group them into similarity clusters. The questionnaire was made available on a website to freshmen physics students at the University Paris Diderot. There were 44 complete responses. Two respondents stood out strongly from the rest, and all the others were divided into three groups with similar misconceptions. Most students do not correctly understand how density, mass, and volume are related, to the point that some of them will consider one of these physical characteristics as independent from the others. Only one participant provided a correct answer to all the questions. Thus, the analysis confirmed that most of the students shared identical misconceptions, and knowledge is lacking in the specific field of density.

68% of the participants went over the 50% rate of correct answers. The average success rate among all participants was 60,31%.

### 5.2 2D game

This experiment was performed on a sample of nine participants, all future physics teachers in the first year studying at the University Paris Diderot. Eight of them had a scientific background, and one of them had a background in literature. Also, we asked them when they last studied the properties of matter: three of them answered "This year during my studies", three others "During my graduate studies", the remaining three said "During my high school studies".

Participants spent an average of 28 min. on the game, with a minimum time of 11min. and a maximum time of 1 hour (see Table 1).

Regarding results of pre- and post-test, the efficiency of participants decreased by 4,27%, but the level of confidence increased by 11,5%: some participants made more mistakes after the experiment than before, but they also gained subjective confidence. This shows that a self-assessed feeling of success is not always the result of successful teaching.

Analyzing answers from tests makes it clear that students still do not correctly interpret how mass, volume, and density are related. Also, a problem arose from the relationship between volume and density in conjunction with a flotation situation. Thus, we see that the concept of density remains challenging to apply to certain practical cases and that the initial misconceptions of students are not easily dispelled.

Despite the inefficacy of 2D games, some students commented that they positively describe such a gaming experience in learning, and their impressions were very good.

**Table 1.** Time for 2D game (unit: minutes).

| Sections | Min t | Max t | Average t |
|---|---|---|---|
| Pre-test | 3,62 | 33,04 | 10,15 |
| Training | 0,51 | 6,37 | 2,44 |
| Scenario 1 | 1,15 | 5,59 | 2,86 |
| Scenario 2 | 0,94 | 3,20 | 1,66 |
| Scenario 3 | 0,84 | 3,45 | 2,22 |
| Scenario Bonus | 0,14 | 1,40 | 0,64 |
| Post-test | 1,98 | 16,65 | 5,31 |
| Total game time | 11,52 | 60,02 | 28,41 |

**Table 2.** Time for 3D game (unit: minutes).

| Sections | Min t | Max t | Average t |
|---|---|---|---|
| Pre-test | 4,54 | 9,92 | 6,62 |
| Training | 0,33 | 2,80 | 1,06 |
| Scenario 1 | 2,76 | 6,99 | 4,34 |
| Scenario 2 | 1,03 | 3,96 | 1,80 |
| Scenario 3 | 2,03 | 3,58 | 2,89 |
| Scenario Bonus | 0,29 | 2,91 | 1,28 |
| Post-test | 3,32 | 7,97 | 4,65 |
| Total game time | 16,23 | 37,30 | 24,07 |

### 5.3 3D game

Because of the prevailing sanitary conditions, only preliminary tests could be conducted on seven participants (Mean age = 23, SD = 2.6) from the laboratory. None of the participants was involved in the design or in the research of the experiment. Almost all of them have a scientific background and play video games every day. Also, 4 of them mentioned that they last studied the properties of matter while in high school and others in middle school.

Playing the density games increased test accuracy from 73,62% to 81,31%, so we can hypothesize that actual 3D manipulations, even with virtual objects, carry more sense and provide more benefits than simple observation in 2D. Similarly, as in the 2D game, the participants increased their confidence by 11,25%. Thus, the actual manipulations were also helpful to bring awareness of some delusions and dispel them, building a correct understanding of the concept of density.

Participants spent 6,62 min. on the pre-test and 4,64 min. on the post-test on average (See Table 2). Analysis answers of tests clearly shows that students have become better at understanding the ratio of density, mass, and volume, but relative flotation remains a more complex concept to grasp.

The average playing time decreased compared to a 2D game. Interestingly, time increased for the Bonus level: Students were more interested because they never tested quicksilver in real life (it is banned from classrooms due to its dangerosity), and the behavior of the cubes surprised them (All objects float in this high-density liquid, and if placed in the middle of it and released, a cube seems to "shoot" out of it).

Also, the salient facts we uncovered were the following:
- General VR scene settings and manipulation procedures were considered satisfactory. Subjects unanimously positively valued all conditions, finding them fun, easy, and they felt confident during the games.
- Most of the participants evaluated sound effects as valuable, giving them additional information and allowing them to focus on a task, and not memorizing which cube is which. Also, 3D sound helped them to get involved in the game.
- Regarding learning, most users established the relationship between the number of points on cubes and density and mass. They were able to apply this in the experiment, but the absence of a similar metaphor for the liquid was sometimes perceived negatively.
- The participants also highly appreciated the proposed learning opportunities by VR (e.g., playing with quicksilver) and found them helpful.

## 6  Conclusions

Students routinely experience difficulties when trying to apply theoretical knowledge in practical situations. They know which formula to use, but they do not develop a deep understanding of the topic of density.

Although it has never been used for the theme of density, VR has repeatedly proven its ecological validity and usefulness in other contexts (such as procedural learning). The results presented in this study showed that immersive serious games could, if not eradicate students' false conceptions on this issue, at least shake them and help look at it from a different viewpoint, allowing them to reconsider their view on the physical phenomenon.

Comparison of 2D and 3D gaming conditions showed the inefficiency of the former and the benefits of immersive games for learning. The 2D game does not provide a sufficient learning experience. On the contrary, the association of VR, serious games, and didactics led to much better results. We have proven that actual manipulation of virtual objects is more beneficial than just observing or interacting with a computer mouse.

It is difficult to force a student to study well without motivation. The students highly appreciated the use of VR for teaching, leading to more motivation. The use of VR associated with traditional teaching models can be a significant impetus for better learning and in-depth understanding of certain physics phenomena.